\documentclass[prx,aps,showpacs,twocolumn,preprintnumbers,
amsmath,amssymb,superscriptaddress,longbibliography]{revtex4-1}
\usepackage[english]{babel}
\usepackage{amsmath,amssymb,amsfonts}
\usepackage{graphicx}
\usepackage[colorlinks=True,linkcolor=red,citecolor=blue,urlcolor=blue]{hyperref}
\usepackage{bookmark}
\usepackage[dvipsnames]{xcolor}
\usepackage{chngcntr}
\usepackage{braket}
\usepackage{nicefrac}
\usepackage{bm}
\usepackage{bbm}
\usepackage{braket}
\usepackage{slashed}
\usepackage[normalem]{ulem}
\usepackage{array}
\newcolumntype{C}{>{$}c<{$}}
\usepackage{chemmacros}

\newcommand{\ve}{\varepsilon}

\newcommand{\bea}{\begin{eqnarray}}
\newcommand{\eea}{\end{eqnarray}}
\newcommand{\be}{\begin{equation}}
\newcommand{\ee}{\end{equation}}

\newcommand{\St}{St\v{r}eda }
\newcommand{\h}{\hat}

\newcommand{\dg}{\dagger}
\newcommand{\hc}{\mbox{h.c.}}

\definecolor{Nathanblue}{rgb}{0.94,0.317,0.5607}

\begin{document}
\bibliographystyle{apsrev4-1}







\title{Energy-Resolved Quantum Geometry from \St Response: \\
Driven-Dissipative Bosonic Lattices and Disordered Systems}

\author{Ana\"is Defossez}
\email{anais.defossez@lkb.ens.fr}
\affiliation{Laboratoire Kastler Brossel, Coll\`ege de France, CNRS, ENS-Universit\'e PSL,
Sorbonne Universit\'e, 11 Place Marcelin Berthelot, 75005 Paris, France
}
\affiliation{International Solvay Institutes, 1050 Brussels, Belgium}
\affiliation{Center for Nonlinear Phenomena and Complex Systems, Universit\'e Libre de Bruxelles, CP 231, Campus Plaine, B-1050 Brussels, Belgium}
\author{Baptiste Bermond}
\affiliation{Laboratoire Kastler Brossel, Coll\`ege de France, CNRS, ENS-Universit\'e PSL,
Sorbonne Universit\'e, 11 Place Marcelin Berthelot, 75005 Paris, France
}
\author{Lucila Peralta Gavensky}
\affiliation{International Solvay Institutes, 1050 Brussels, Belgium}
\affiliation{Center for Nonlinear Phenomena and Complex Systems, Universit\'e Libre de Bruxelles, CP 231, Campus Plaine, B-1050 Brussels, Belgium}
\author{Nathan Goldman}
\email{nathan.goldman@lkb.ens.fr}
\affiliation{Laboratoire Kastler Brossel, Coll\`ege de France, CNRS, ENS-Universit\'e PSL,
Sorbonne Universit\'e, 11 Place Marcelin Berthelot, 75005 Paris, France
}
\affiliation{International Solvay Institutes, 1050 Brussels, Belgium}
\affiliation{Center for Nonlinear Phenomena and Complex Systems, Universit\'e Libre de Bruxelles, CP 231, Campus Plaine, B-1050 Brussels, Belgium}

\begin{abstract}
The \St formula links the Hall conductivity of an insulator to the magnetic‑field response of its particle density, providing a local and universal probe of the topological Chern number. Beyond this quantized response, an energy‑resolved \St marker can be defined from the magnetic response of the density of states, revealing detailed features of the quantum geometry of Bloch bands. We show that driven–dissipative bosonic lattices provide direct access to both the integrated and energy‑resolved \St responses. Our scheme uses controlled pumping with uniform strength and random phases across the lattice, together with uniform loss, to yield a Lorentzian filter of eigenmode occupations. For generic dispersive bands, this enables reconstruction of a coarse‑grained energy‑resolved Středa response, establishing these platforms as versatile probes of anomalous spectral flow and energy‑resolved quantum geometry. As a striking application, we show that this marker elucidates the fate of topological bands under strong disorder, capturing the quantum‑geometric structure underlying topological Anderson insulators.
\end{abstract}

\date{\today}
\maketitle

\textit{Introduction.} 
Chern insulators are formally distinguished from trivial ones by a topological invariant defined for gapped bulk bands:~the Chern number~\cite{Thouless_1982}. In the presence of edges, this invariant acquires a direct physical meaning through the emergence of gapless boundary modes, whose number is fixed by the bulk–edge correspondence~\cite{Halperin_1982,Hatsugai1993}.
While the Chern numbers of the occupied Bloch bands dictate the quantized Hall conductivity~\cite{Thouless_1982, Avron_1985, Niu_1985}, they also determine the response of the particle density ($n$) to a uniform magnetic field ($B$), a relation known as the Středa formula~\cite{Streda_1982, Widom_1982, Streda_Smrcka_1983}, 
\begin{equation}
    \Phi_0 \, \frac{\partial n}{\partial B} =\frac{\sigma_H}{\sigma_0}= \sum_{\beta \in {\rm occ}} \mathcal{C}_{\beta}. \label{eq: streda thermod}
\end{equation} 
Here $\mathcal{C}_{\beta}$ denote the Chern numbers of the occupied bands; $\sigma_0 = e^2/h$ is the conductance quantum and $\Phi_0 = hc/e$ the flux quantum.

The density response in Eq.~\eqref{eq: streda thermod} can be understood as a modification of the density of states (DOS) induced by the magnetic perturbation via the Berry curvature~\cite{Niuetal2005, Bliokh_2006, Duval_2006}. 
In fact, an energy-resolved St\v{r}eda formula, entirely expressed in terms of the Bloch band density of states, has been recently derived for non-interacting systems \cite{PeraltaGavensky_2025},
\begin{align}
    \Phi_0 \frac{\partial \rho(\omega)}{\partial B}\Big|_{B=0} &= \int_{\mathrm{BZ}} \frac{d^2k}{2 \pi} \sum_{\beta} \left[ \mathcal F^\beta_{xy}(k) \delta(\omega - \varepsilon_{\beta k}) \right. \notag \\ 
    &\ \left. + \,  \frac{\Phi_0}{2\pi} m_z^\beta(k)\frac{\partial}{\partial \omega} \delta(\omega - \varepsilon_{\beta k}) \right] \, , \label{eq: energy resolved Streda}
\end{align}
where $\rho(\omega)$ denotes the DOS of the system, and $\mathcal F^\beta_{xy}(k)$ and $m_z^\beta(k)$ stand for, respectively, the Berry curvature and intrinsic orbital magnetic moment of the $\beta$th Bloch band defined in $k$-space~\cite{Xiao_2010}. The quantized response in Eq.~\eqref{eq: streda thermod} then directly follows from Eq.~\eqref{eq: energy resolved Streda} upon a complete occupation of the Bloch bands. 

\begin{figure}[h]
    \centering
    \includegraphics[width=\columnwidth]{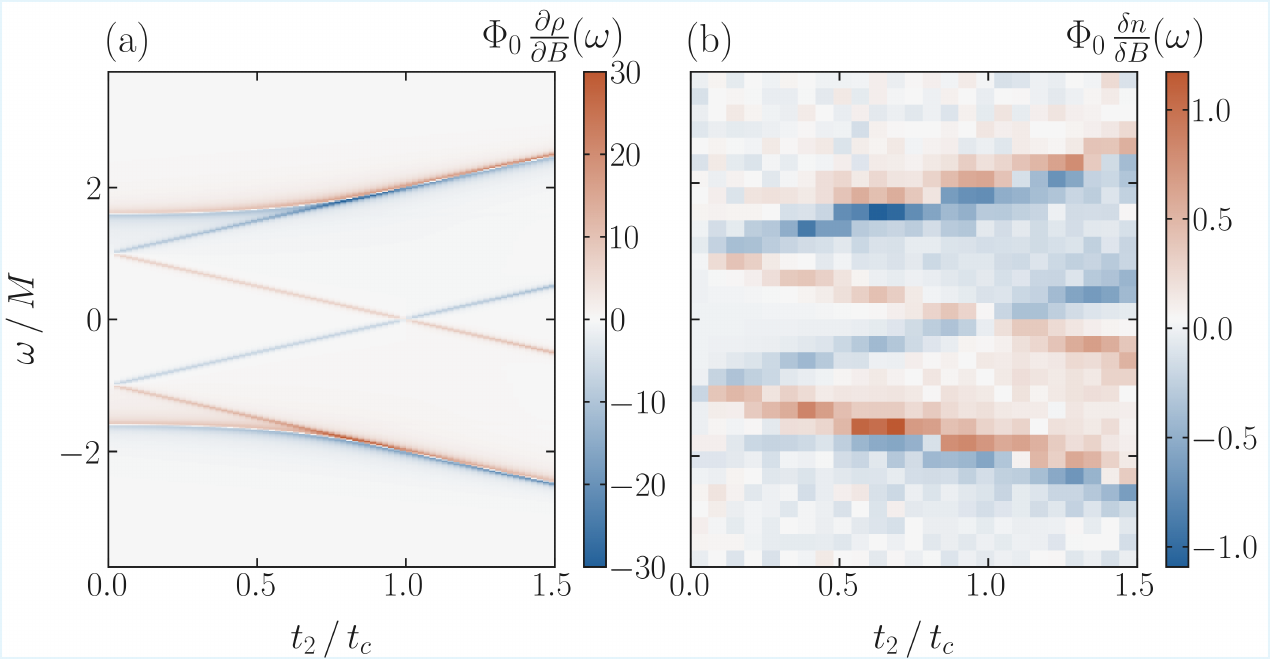}
    \caption{(a) Energy-resolved \St response [Eq.~\eqref{eq: energy resolved Streda}] across a topological transition ($t_2\!=\!t_c$) in the Haldane model with periodic boundaries. (b) The corresponding coarse‑grained density response obtained from the driven-dissipative protocol, considering the Haldane model with open boundaries. The magnetic perturbation is set to $\delta \alpha = 1/160$, and the photon density is averaged over $D = 60$ realizations. The parameters of the model are set to $M = 0.8 \, t_1$ and $\phi = \pi/2$.}
    \label{fig: colormap energy res streda}
\end{figure}

Importantly, the result in Eq.~\eqref{eq: energy resolved Streda} reveals an energy-resolved marker of quantum geometry, allowing one to pinpoint the ‘hot spots’ of the Berry curvature~\cite{Yang2011,Qiao2016}. Singular features of the density of states, i.e.~van Hove singularities~\cite{van1953occurrence}, also contribute to this energy-resolved Středa response. 
The quantum-geometric fine structure revealed by the energy-resolved Středa marker is remarkably rich, exposing subtle features of topological bands under external perturbations. Its diagnostic power becomes especially striking across topological phase transitions, as illustrated in Fig.~\ref{fig: colormap energy res streda}(a). An especially appealing -- and unexplored -- aspect of this Středa marker is its ability to track the evolution of topological bands in the presence of strong disorder~\cite{Prodan2011}, a capability we demonstrate explicitly toward the end of this Letter.

 While the quantized density response in Eq.~\eqref{eq: streda thermod} relies on the complete filling of Bloch bands, the marker in Eq.~\eqref{eq: energy resolved Streda} does not, as it is a property of the single-particle band structure. This makes it possible to explore the energy-resolved quantum geometry of Bloch systems in a wide range of settings, including bosonic setups. In this Letter, we explore this possibility by proposing that the energy-resolved \St marker can be measured in driven-dissipative bosonic platforms, such as photonic lattices~\cite{lu2014topological, khanikaev2017two, ozawa2019topological}. In particular, we demonstrate that selectively populating states within a targeted energy window using driven-dissipative (DD) protocols~\cite{Carusotto_Ciuti_2013, Ozawa_2014, Schiroetal_2025, Bermond_2025} provides a direct route to access the energy-resolved \St marker in Eq.~\eqref{eq: energy resolved Streda}. Concretely, we derive the steady-state solutions of a Lindblad master equation incorporating both dissipation and an external driving pump, specifically engineered to stabilize the targeted states. Building on this driven–dissipative framework and using the Haldane model as an illustrative example, we show that, for generic dispersive bands, the protocol provides access to a ``coarse‑grained" energy‑resolved \St response. Figure~\ref{fig: colormap energy res streda}(b) illustrates the energy-resolved \St response obtained from this DD protocol, across a topological transition. 

The remainder of the Letter is organized as follows. We first introduce the driven‑dissipative protocol that enables energy‑filtered population of Bloch states in bosonic systems. We then apply this scheme to a bosonic implementation of the Haldane model, from which we extract the energy‑resolved \St marker and examine the quantum‑geometric fine structure across a topological transition. Finally, we demonstrate the diagnostic power of this \St marker in strongly disordered topological systems, linking this quantum‑geometric signature to the persistence -- or revival -- of extended states.


\textit{Driven-dissipative framework.} 
Photonic systems offer fine control over photon injection and leakage~\cite{Carusotto_Ciuti_2013,ozawa2019topological}. We introduce a protocol compatible with such platforms, in which a coherent external pump at frequency $\omega_0$
 injects photonic excitations across the lattice, in combination with spatially uniform dissipation. As we show below, this driving scheme can stabilize non‑trivial, out‑of‑equilibrium steady states, suitable for \St-type measurements.

The total Hamiltonian of the system is written in the form $\h H = \h H_{cav} + \h H_{pump}$, including both the cavity modes and the external pump, respectively. 
In the following, we denote by $\h c^{(\dg)}$ the photon creation/annihilation operators, and consider a linear-optics regime where photon–photon interactions can be neglected~\cite{Carusotto_Ciuti_2013}. The cavity Hamiltonian is then quadratic, and can be written as $\h H_{cav} = \sum_{\mu \nu} \h c^\dg_\mu H_{\mu \nu} \h c_\nu$, with $H_{\mu \nu}$ the matrix elements in the basis spanned by $\h c_\mu$. The coherent external pump is captured by $\h H_p = \sum_i f_i\, e^{-i\omega_0 t} \h c^\dg_i + \hc$ with $f_i$ the pump strength at site $i$.
To engineer a uniform photon distribution, both spatially and around a targeted energy window of the spectrum, we consider a pumping with uniform amplitude $(f)$ and random phases across the lattice, defined as 
\begin{align}
    f_i = f e^{i\theta_i} \, , \label{eq: uniform pump random}
\end{align} where $\theta_i \in [0,2\pi[$ are independent and uniformly distributed random variables.
Assuming Markovian radiative coupling, dissipation is included using a Lindblad formalism with a loss rate chosen to be uniform on the lattice, $\gamma_i = \gamma, \,  \forall i$.

We then define the photon mean fields in the rotating frame induced by the pump, $c_i \equiv e^{i \omega_0 t} \langle \h c_i (t) \rangle$. 
In the zero-temperature limit, the Lindblad-type Heisenberg equation for the mean fields $\langle \h c_i (t) \rangle$ yields an equation of motion for the amplitudes $c_i$~\cite{Ozawa_2014,Price_COM}
\begin{align}
    -i \partial_t c_i = \left( (\omega_0 + i\gamma) \delta_{ij} - H_{ij} \right) \, c_j - f_i \, , \label{eq: reduced Lindblad}
\end{align}
where implicit summation over repeated indices is assumed. For steady-state solutions $\partial_t c_i = 0$, this equation reduces to a linear system.\\
For the random pump defined in Eq.~\eqref{eq: uniform pump random}, and writing the cavity Hamiltonian with open boundaries in diagonal form, $\hat H = \sum_\lambda \varepsilon_\lambda \hat c^\dagger_\lambda \hat c_\lambda$, one can show that the expectation value of the steady-state photon occupation of each eigenmode follows a Lorentzian distribution [Appendix]
\begin{align}
\mathbb E[|c_\lambda|^2] = \frac{|f|^2}{(\omega_0 - \varepsilon_\lambda)^2 + \gamma^2} \, . \label{eq: exp value eigenstates occupation}
\end{align}
Repeating over random spatial phases destroys coherent interferences between lattice sites, so that the drive behaves effectively as an incoherent sum of local pumps: each eigenmode is thus populated independently with a Lorentzian weight of width determined by the dissipation $\gamma$ and centered at $\omega_0$. A similar configuration can be realized experimentally using an incoherent pump, as is commonly employed in photoluminescence measurements~\cite{jamadi2020direct}.

In this work, we propose to monitor the \St response of the bulk photon density within this DD protocol. Here, we define the photon density as the average light intensity within a region of area $A$ located in the bulk of the lattice,
\begin{align}
    n(\omega_0) = \frac{1}{A} \sum_{i \in A} |c_i|^2\, .\label{eq: photon bulk density}
\end{align}
The photon amplitudes $c_i$ correspond to the steady-state solutions of the DD equation~\eqref{eq: reduced Lindblad}, for a given pump frequency $\omega_0$ and loss $\gamma$. Given the mean occupations in Eq.~\eqref{eq: exp value eigenstates occupation}, the expectation value of the photon density is determined by 
\begin{align}
    \mathbb E[n(\omega_0)] &= \int \rho_A(\omega) \frac{|f|^2}{(\omega_0 - \omega)^2 + \gamma^2} d\omega \, , \label{eq: exp value photon bulk intensity}
\end{align}
where 
\begin{align}
    \rho_A(\omega) = \frac{1}{A} \mathrm{Tr} [\h P_A \, \delta(\omega - \h H)] \label{eq: bulk proj DOS}
\end{align}
is the bulk DOS with $\h P_A$ a projector onto the considered subregion. The result in Eq.~\eqref{eq: exp value photon bulk intensity} indicates that the light intensity $n$ can be used to monitor the bulk DOS $\rho_A (\omega)$, within an energy-window determined by the pump frequency $\omega_0$ and the loss rate $\gamma$. In practice, the DOS can also be monitored through photoluminescence measurements~\cite{jamadi2020direct}.

While photonic lattices offer a convenient platform for implementing this scheme, it is in fact applicable to generic bosonic systems, provided the mean-field regime is valid and suitable driven-dissipative protocols are available; see the proposal of Ref.~\cite{Bermond_2025} for magnonic spin systems. 

\textit{Measuring the energy-resolved St\v{r}eda response.} 
The energy-resolved \St response in Eq.~\eqref{eq: energy resolved Streda} reflects the modification of the bulk DOS $\rho (\omega)$ under an orbital magnetic perturbation. In the present DD framework, one proposes to access this \St marker by monitoring the response of the bulk light intensity [Eqs.~\eqref{eq: reduced Lindblad} and \eqref{eq: photon bulk density}] to a magnetic perturbation,
\begin{equation}
\Phi_0 \, \Bigl \langle \frac{\delta n}{\delta B} \Bigr \rangle_D \approx \Phi_0 \, \frac{1}{\delta B} \langle n (\delta B) - n \rangle_D ,\label{eq_finite_difference}
\end{equation}
where $n (\delta B)$ denotes the modified bulk density, and $\langle \cdot \rangle_D$ is the average over $D$ realizations. In the photonics context, the ``magnetic" perturbation $\delta B$ corresponds to a synthetic gauge field, which can be engineered through various methods~\cite{lim2017electrically,ozawa2019topological}. Following Eqs.~\eqref{eq: exp value photon bulk intensity}-\eqref{eq_finite_difference}, and considering a sufficiently large number of realizations $D$, one obtains a ``coarse-grained" energy-resolved \St marker  
\begin{align}
    \Phi_0 \, \Bigl \langle \frac{\delta n(\omega_0)}{\delta B} \Bigr \rangle_D \approx \Phi_0 \int \frac{\delta \rho_A(\omega)}{\delta B} \frac{|f|^2}{(\omega - \omega_0)^2 + \gamma^2} d\omega \, . \label{eq: convolved Streda}
\end{align}
This corresponds to the energy‑resolved \St response in Eq.~\eqref{eq: energy resolved Streda}, convolved with a Lorentzian kernel
that captures the eigenmode occupations set by the pump parameters and the loss rate [Eq.~\eqref{eq: exp value eigenstates occupation}]. The resulting observable in Eq.~\eqref{eq: convolved Streda} therefore provides a finite-resolution response centered at the pump frequency $\omega_0$, with resolution set by the loss rate $\gamma$.  For a dispersive Bloch band of width $\Delta_w$, and in the regime of small loss $\gamma\!\ll\!\Delta_w$, the energy‑resolved \St response in Eq.~\eqref{eq: energy resolved Streda} can thus be reconstructed by scanning the pump frequency $\omega_0$ in discrete steps of size 
$\delta \omega_0 \propto \gamma$. The Chern number of the band can eventually be obtained by summing the coarse-grained \St response in Eq.~\eqref{eq: convolved Streda} over the entire band, reaching a quantized value in the limit $\gamma \to 0$; we note that this integrated \St response also includes a filling factor established by the ratio $|f|/\gamma$; see Appendix.


\begin{figure}[b]
    \centering
    \includegraphics[width=\columnwidth]{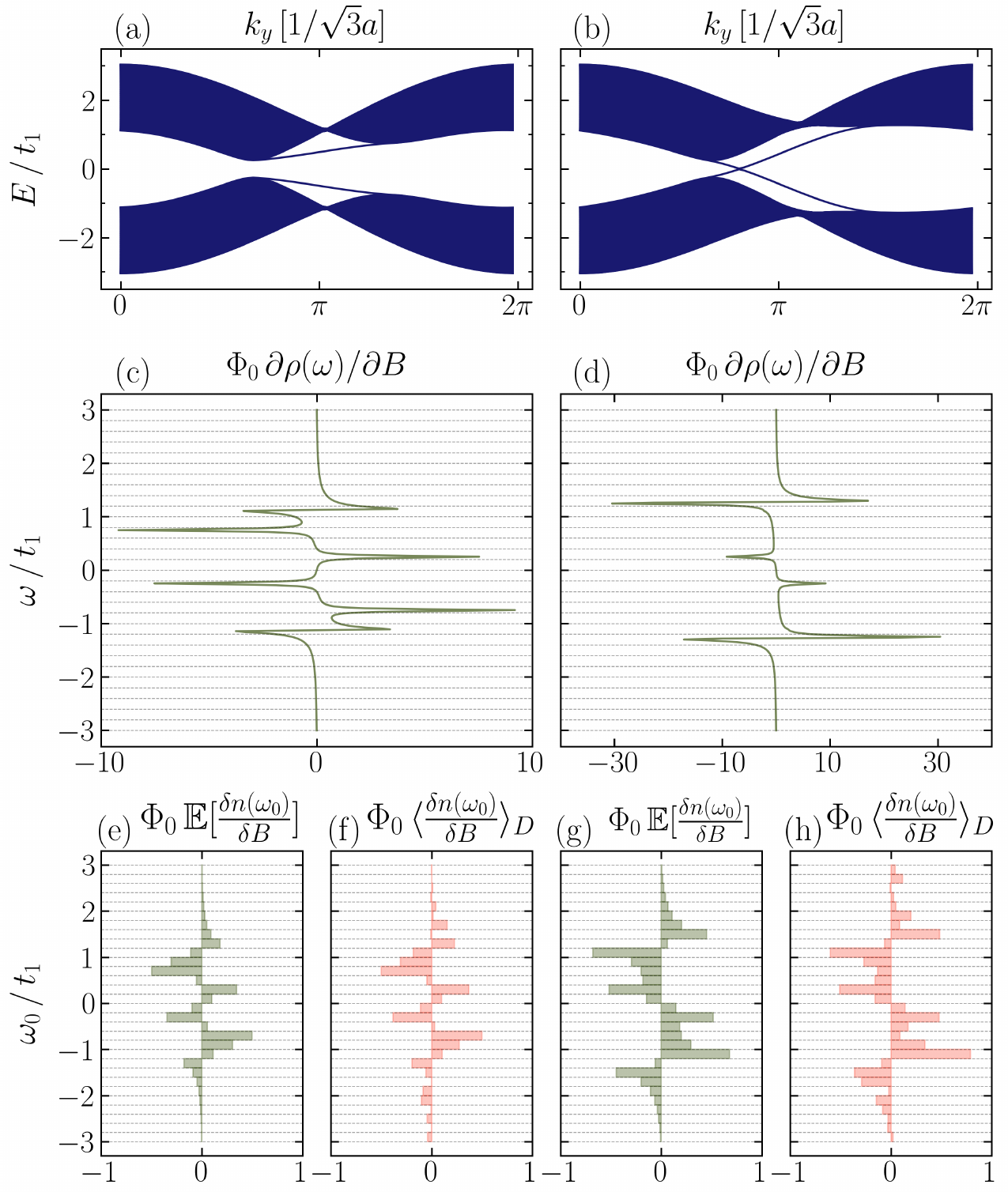}
    \caption{(a)-(b) Energy spectrum of the Haldane model in a cylindrical geometry ($N_x = 75$ cells) as a function of quasimomentum $k_y$,  shown for the trivial regime ($t_2=t_c/2$) and the topological regime ($t_2=1.5\,t_c$), with $\phi=\pi/2$ and $M=0.5 \,  t_1$. (c)-(d) Analytical energy-resolved Středa response [Eq.~\eqref{eq: energy resolved Streda}] for the same parameters, computed with periodic boundary conditions. (e)-(h) Coarse-grained energy-resolved St\v{r}eda response obtained in a finite system ($55 \times 55$ unit cells), with a magnetic perturbation $\delta\alpha=1/160$, loss rate $\gamma/t_1 =0.1$ and drive amplitude $f = \gamma$. (f),(h) Photon density response $\langle \Phi_0\,\delta n (\omega_0) /\delta B \rangle_D$ within a  bulk subregion of size $45 \times 45$, extracted from the steady-state solutions of the DD master equation [Eq.~\eqref{eq: photon bulk density}] and averaged over $D=60$ realizations. The response is displayed using $N_\omega=30$ energy-resolved pixels of width $2\gamma$ centered at frequencies $\omega_0$ (dashed lines). (e),(g)  Corresponding analytical prediction from Eq.~\eqref{eq: convolved Streda}.}
    \label{fig: energy resolved trivial and non trivial}
\end{figure}
In the flat‑band regime, the energy resolution becomes irrelevant. When the loss rate lies between the band width and the band gap, $\Delta_w \ll \gamma \ll \Delta_g$, and the pump frequency $\omega_0$ is resonant with a chosen Bloch band, the drive uniformly populates that band with a constant filling factor~\cite{Ozawa_2014}. In this limit, the occupations in Eq.~\eqref{eq: exp value eigenstates occupation} reduce to the uniform value $\mathbb E[|\ve_\lambda|^2] \approx (|f|/\gamma)^2$. Consequently, the density response in Eq.~\eqref{eq: convolved Streda} directly yields the Chern number of the targeted Bloch band [Eq.~\eqref{eq: streda thermod}], quantized up to the overall filling factor; see Appendix.
\textit{Numerical validation.} We numerically implement the protocol on the Haldane model \cite{Haldane_1988} defined on a honeycomb lattice. The nearest-neighbor hopping amplitude is denoted by $t_1$, while staggered on-site energies $\pm M$ are assigned to the two sublattices. Time reversal symmetry is broken by complex second-neighbor hoppings $t_2 e^{\pm i \phi}$. The model undergoes a topological phase transition at a critical hopping amplitude $t_2 = t_c \equiv \left| \frac{M}{3 \sqrt{3} \sin(\phi)} \right|$.
The magnetic perturbation $\delta B$ is introduced through Peierls phases in the hopping matrix elements and expressed in terms of the dimensionless flux $\delta \alpha = A_{\mathrm{cell}} \, \delta B/\Phi_0$, with $A_{\mathrm{cell}}$ the area of a unit cell of the honeycomb lattice. Figures~\ref{fig: energy resolved trivial and non trivial}(a) and (b) show the energy spectra of the model in the trivial and non-trivial regimes, respectively, computed in a cylinder  geometry to visualize the edge modes. Figures~\ref{fig: energy resolved trivial and non trivial}(c) and (d) show the corresponding analytical energy-resolved St\v{r}eda response [Eq.~\eqref{eq: energy resolved Streda}] obtained under periodic boundary conditions. 

We next apply the protocol for these two distinct regimes,  considering a finite system with open boundaries  consisting of $55 \times 55$ unit cells. The magnetic perturbation is set to $\delta \alpha = 1/160$, with loss rate $\gamma / t_1 = 0.1$, and the protocol is performed for $N_\omega=30$ pump frequencies $\omega_0$ scanning the full spectrum. In Figs.~\ref{fig: energy resolved trivial and non trivial}(e)-(h), the \St response is displayed for each pumping frequency using an energy-resolved grid of width $2\gamma$ centered  at $\omega_0$. The response $\Phi_0 \langle \delta n / \delta B\rangle_D$ is obtained from the steady-state solutions of the DD master equation [Eq.~\eqref{eq: photon bulk density}], averaged over $D=60$ realizations, which we benchmark with the expectation value predicted for an infinite number of realizations [Eq.~\eqref{eq: convolved Streda}].
We find good agreement between the St\v{r}eda response obtained from the DD protocol [Figs.~\ref{fig: energy resolved trivial and non trivial}(f) and (h)] and the corresponding theoretical predictions [Figs.~\ref{fig: energy resolved trivial and non trivial}(e) and (g)].  In particular, the  protocol reliably captures both the Berry-curvature hot spots and the van Hove singularities  of the spectrum, which are directly encoded in the analytical energy-resolved St\v{r}eda expression shown in Figs.~\ref{fig: energy resolved trivial and non trivial}(c)-(d).

The complete exploration of the energy-resolved marker across the topological transition is shown in Fig.~\ref{fig: colormap energy res streda}. Panel (a) displays the analytical energy-resolved \St response as a function of $t_2/t_c$ across the transition \cite{PeraltaGavensky_2025}, while Fig.~\ref{fig: colormap energy res streda}(b)  shows the corresponding results obtained from the DD protocol (averaged over $D=60$ realizations). We used the same system size and bulk subregion, as well as the same loss rate $\gamma$ and drive amplitude $f$ as in Fig.~\ref{fig: energy resolved trivial and non trivial}. The two panels are in excellent agreement, confirming that the protocol provides a reliable energy-resolved probe of quantum geometry.

Finally, we note that probing the energy-resolved St\v{r}eda response in a finite system through this protocol involves an interplay between the magnetic perturbation, the strength of the loss rate, and the intrinsic fluctuations of the protocol, which are set by the number of realizations. These parameters are therefore carefully adjusted in order to clearly resolve the desired physical features. A more detailed analysis is provided in the Appendix. \\

\begin{figure}[h]
    \centering
    \includegraphics[width=\columnwidth]{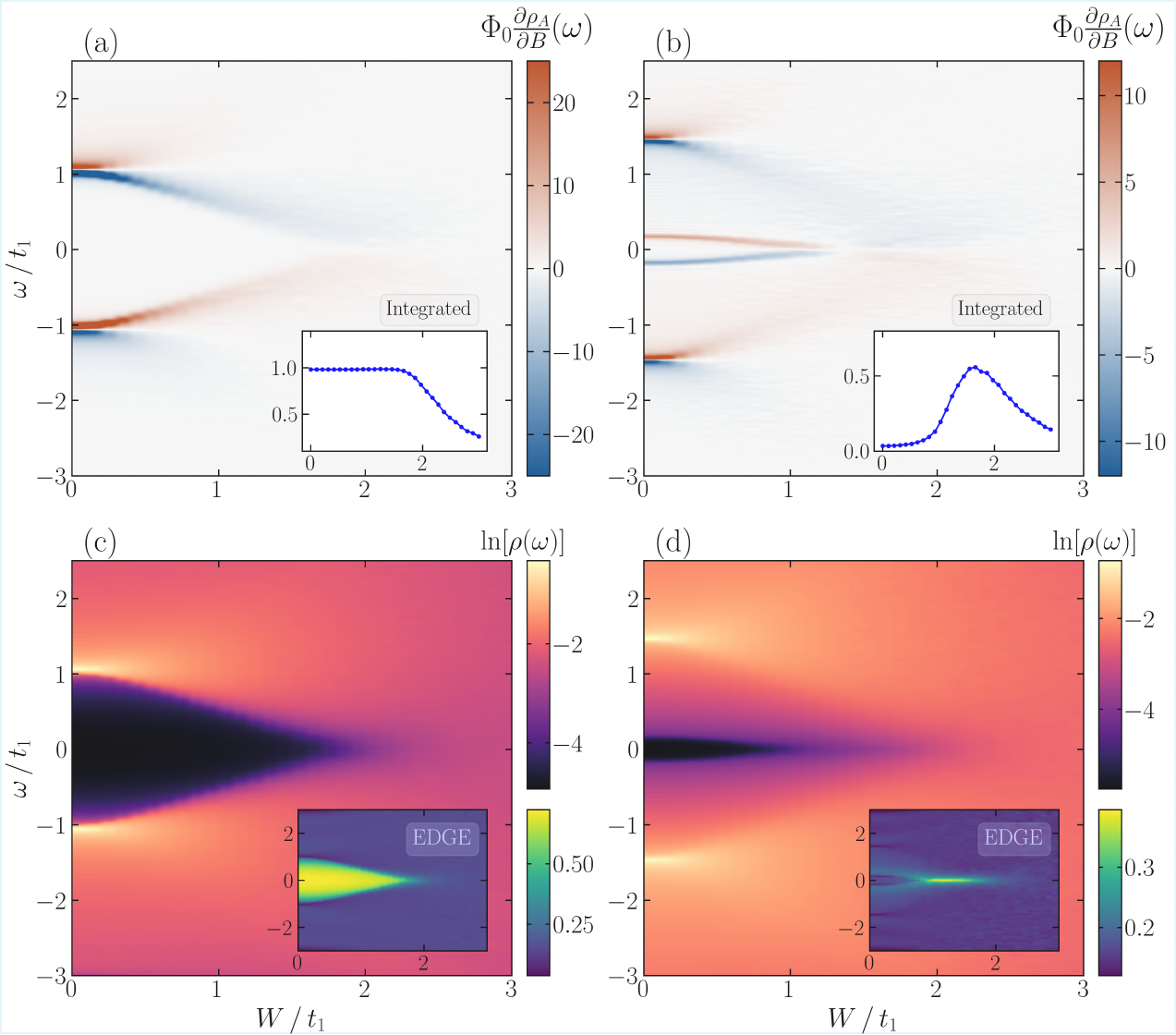}
    \caption{(a)-(b) Energy-resolved \St response evaluated as a finite difference of the bulk DOS $\rho_A(\omega)$ [Eq.~\eqref{eq: bulk proj DOS}] of the Haldane model of size $51 \times 51$ as a function of disorder strength $W$. The bulk subregion $A$ is obtained by removing two layers from the boundary of the sample. The magnetic perturbation is set to $\delta \alpha = 1/100$ (a) and $\delta \alpha = 1/300$ (b). For a clear visibility, the colorbars are clipped at $\Phi_0 \delta \rho_A (\omega)/ \delta B \in [-25, 25]$ (a) and $[-12,12]$ (b). In each panel (a) and (b), we provide an inset with the associated integrated response up to $\omega = 0$ as a function of disorder. 
    The system is initialized in a non-trivial regime with $M/t_1 = 0$ and $t_2/t_1 = 0.2$ (a), and in a trivial regime with $M/t_1 = 0.8$ and $t_2 = 0.8 \, t_c$ (b).
    Panels (c) and (d) respectively display the DOS $\rho(\omega)$ at $B = 0$, and the corresponding edge-projected fraction of the DOS as insets, for a two-layer thick edge. The (bulk) density of states is computed through a Lorentzian approximation with a broadening $\eta = 0.02 \, t_1$.}
    \label{fig: streda disorder}
\end{figure}
\textit{Energy-resolved \St response in disordered systems.}  
We now show that the energy-resolved \St response is particularly well suited to reveal quantum-geometric and topological properties in strongly disordered systems, i.e.~when Bloch-band structures are strongly altered. In particular, we consider gapped disordered systems, where topological properties can still manifest, through the survival (or revival) of extended states~\cite{Halperin_1982,Laughlin_1984,Prodan2011,Li2009,Groth2009,Zhang_2012, Liu_2023}. According to Laughlin's argument~\cite{Laughlin_1984}, these extended states carry the spectral flow under a magnetic perturbation, and thus maintain the topological nature of the system until they annihilate with extended states of opposite character. As such, these extended states are expected to leave a clear fingerprint in the \St response, making the latter a natural probe of topology in disordered systems. 

To investigate this effect, we add on‑site disorder to the Haldane model, with amplitudes uniformly distributed in $[-W, W]$. We then study the energy-resolved \St~response -- computed by directly evaluating the magnetic response of the bulk density of states in Eq.~\eqref{eq: bulk proj DOS} -- as a function of the disorder strength. Our analysis is performed for both the trivial and topologically non-trivial regimes of the clean Haldane model ($W\!=\!0$).

In the non-trivial regime shown in Fig.~\ref{fig: streda disorder}(a), the \St~response exhibits a clear signal concentrated at energies that track the band edges as disorder increases. This behavior can be interpreted as a redistribution of Berry-curvature hot spots induced by disorder, and is consistent with Laughlin's levitation scenario for the underlying extended states, which remain robust up to a critical disorder strength $W_c$ at which the gap closes; see the DOS in Fig.~\ref{fig: streda disorder}(c). To validate this interpretation, we compute an ``edge-state marker'', obtained by projecting the density of states onto a narrow region along the system's boundary. The result, shown in the inset of Fig.~\ref{fig: streda disorder}(c), confirms that robust edge modes persist throughout the topological regime identified by the \St response. For completeness, we also present the integrated \St~response (ISR) -- obtained by integrating the energy-resolved signal up to the center of the spectrum (the spectral gap) -- in the inset of Fig.~\ref{fig: streda disorder}(a). The quantized value of the ISR provides a clear hallmark of the gapped topological regime. Beyond $W_c$, this quantization breaks down as Berry-curvature hot spots approach each other in energy and their contributions progressively cancel; see main panel (a). Altogether, these results demonstrate how the energy-resolved \St response reveals the microscopic mechanisms governing both the robustness and the eventual destruction of the topological regime under strong disorder.

A richer phenomenology emerges when starting from the trivial regime of the Haldane model, while still breaking time-reversal symmetry; see Figs.~\ref{fig: streda disorder}(b) and (d). In this configuration, a non-trivial fine structure remains visible in the energy-resolved \St~response. In particular, for each band, one observes pairs of Berry-curvature hot spots of opposite sign, which can be naturally associated with contributions originating from the two inequivalent valleys of the underlying band structure. As disorder increases, the contribution from one valley crosses zero energy and undergoes a sign inversion, while the other one shifts toward zero energy from higher energies. This produces a finite window of disorder in which the Berry curvature structure around $\omega\!=\!0$ is inverted, before being ultimately suppressed at higher disorder. Within this intermediate regime, edge states emerge [inset of Fig.~\ref{fig: streda disorder}(d)], and the ISR [inset of Fig.~\ref{fig: streda disorder}(b)]  increases correspondingly. This is consistent with the opening of a mobility gap and the onset of a topological Anderson insulator (TAI) regime~\cite{Assuncao_2024, Kovalenka_2026}. 
To corroborate this interpretation, we benchmark our results with the local Chern marker [Appendix], which is expected to be comparable to the ISR in Chern insulators~\cite{Repse_2026}. Together with a finite-size scaling analysis, this comparison reveals that the deviation of the ISR signal from a quantized value observed in Fig.~\ref{fig: streda disorder}(b) can be attributed to a finite-size effect. Indeed, for a very small mobility gap, edge states can extend deep into the bulk and affect the marker.

This analysis establishes the energy-resolved \St~response as a powerful probe of quantum-geometrical fine structures across TAI-type transitions. A more complete characterization could be achieved by employing advanced numerical approaches capable of accessing significantly larger system sizes -- thereby reducing edge-related contributions~\cite{Antao_2026} -- as well as through complementary localization analyses~\cite{Zhang_2012}.\\

\textit{Conclusion and perspectives.}
This work establishes driven‑dissipative protocols as a powerful and experimentally accessible framework for probing Středa responses in bosonic systems, enabling the direct measurement of quantized topological invariants and energy‑resolved geometric markers. We further demonstrate that the energy‑resolved Středa marker is a particularly sensitive probe of disorder‑induced quantum geometry and topology, yielding a clear signature of transitions into TAI regimes. The present framework opens several promising directions for future exploration. Extending the scheme to systems subject to a pseudo‑magnetic field -- readily engineered through strain in photonic lattices~\cite{ozawa2019topological} -- could enable practical probes of the Středa response in platforms realizing the valley Hall effect~\cite{Jamotte_2023,noh2018observation,dong2017valley,xue2021topological,liu2021valley}. Further opportunities arise in Floquet bosonic systems~\cite{reichl2014floquet,wintersperger2020realization,eckardt2017colloquium,hesse2025probing}, where the Středa response may provide access to Floquet topological invariants and reveal anomalous dynamical phases~\cite{PeraltaGavensky_2025,gavensky2025quantized,Liu2020},  as well as amorphous~\cite{rechtsman2011amorphou,yang2019topological,tran2015topological} and quasicrystal~\cite{tran2015topological,Segev_QC} systems, where the local Středa response is particularly appealing. Beyond photonics, analogous driven-dissipative strategies could be developed for magnonic platforms~\cite{Bermond_2025,mcclarty2022topological,zhuo2025topological}, broadening the scope of practical implementations.  Another promising route concerns cold atoms in optical lattices, where bosonic particles can be loaded in the band structure in an energy-resolved way~\cite{elevator_Botao}, and the Středa response monitored at the single-site level~\cite{Repellin_box,leonard2023realization}. It would also be valuable to understand how nonlinearities, such as those described by the nonlinear Schrödinger equation~\cite{Carusotto_Ciuti_2013}, affect the Středa response of topological Bloch bands. Indeed, while Eq.~\eqref{eq: energy resolved Streda} is valid for non-interacting particles, the magnetic response of the DOS can be relevant in interacting (possibly strongly-correlated) settings~\cite{peralta2023connecting}. Finally, exploring the quantum regime, where driven-dissipative protocols might enable the preparation of correlated topological states~\cite{Kapit_PRX,ozawa2019topological}, represents an essential step toward assessing the robustness and fundamental limits of the Středa response in genuinely quantum out-of-equilibrium settings.\\


\paragraph*{Acknowledgments} We thank Philippe St-Jean, Sylvain Ravets and Jacqueline Bloch for their insightful comments on our manuscript. We also acknowledge discussions with Marco Schiro, Ivan Amelio and Iacopo Carusotto. This research was financially supported by the ERC Grant LATIS, the FRS-FNRS (Belgium), the EOS project CHEQS, and the Fondation ULB. This work also received support from the French government, managed by the National Research Agency, under the France 2030 program, reference ANR‑23‑PETQ‑0002. L.P.G. acknowledges support provided by the FRS-FNRS Belgium and the L'Or\'eal-UNESCO for Women in Science Programme.

\bibliography{mibib}

\vspace{1cm}

\begin{center}
{\bf Appendix}
\end{center}

\appendix

\vspace{-1.5cm}

\section{Steady-state photon occupation distribution from the random pumping}

We provide here the derivation of Eq.~\eqref{eq: exp value eigenstates occupation}. Our starting point is the Lindblad-type Heisenberg equation of motion for an operator $\h A$
\begin{align}
    i \frac{d\langle \h A \rangle}{dt} = \langle [\h A, \h H]\rangle + i \sum_\sigma \gamma_\sigma \Bigl \langle 2 \, \h L^\dg_\sigma \h A \h L_\sigma -  \left\{ \h A, \h L^\dg_\sigma \h L_\sigma \right\} \Bigr \rangle \, ,
\end{align}
where $\h L_\sigma$ are the jump operators that account for the dissipative channels \cite{Schiroetal_2025}. Taking $\h L_\sigma = \h c_i$, one then obtains the equations of motion for the fields $c_i = e^{i \omega_0 t} \langle \h c_i(t) \rangle$ Eq.~\eqref{eq: reduced Lindblad}. 
For the uniform pumping drive with random phases across the lattice defined in Eq.~\eqref{eq: uniform pump random}, the steady-state solutions result in $M_{ij} \, c_j= f_i$, where $M_{ij} = (\omega_0 + i \gamma) \delta_{ij} - H_{ij}$ and $f_i$ denotes the pumping field defined site by site as $f_i = f e^{i \theta_i}$. \\ 

Let $P$ denote the unitary matrix that diagonalizes the Hamiltonian, such that $H^D = P^\dg H P$, where $H^D$ is diagonal. We denote by $e_\lambda$ the normalized eigenvector of $H$ with eigenvalue $\varepsilon_\lambda$, so that $P_{i\lambda} = (e_\lambda)_i$. The components of the photon field in the eigenmode basis are then given by $c_\lambda = P^\dg_{\lambda i} c_i$.
Using $c_i = M^{-1}_{ij}f_j$, we obtain
\begin{align}
|c_\lambda|^2 &= \sum_{ij} c^\dg_i P_{i \lambda} P^\dg_{\lambda j} c_j \\
&= |f|^2 \sum_{ij} \sum_{kl} e^{-i(\theta_l - \theta_k)} (M^{-1})^\dg_{li} P_{i \lambda} P^\dg_{\lambda j} M^{-1}_{jk} \notag \\
\implies \mathbb E[|c_\lambda|^2] &= |f|^2 \sum_{ij} \sum_k (M^{-1})^\dg_{ki} P_{i \lambda} P^\dg_{\lambda j} M^{-1}_{jk} \label{eq: exp value developement} \\
&= \frac{|f|^2}{(\omega_0 - \varepsilon_\lambda)^2 + \gamma^2} \sum_k (e_\lambda)_k (e^*_\lambda)_k \notag \\
&= \frac{|f|^2}{(\omega_0 - \varepsilon_\lambda)^2 + \gamma^2} \notag \, ,
\end{align}
where we used the independence of the phases and their uniform distribution, $\mathbb E[e^{i(\theta_k - \theta_l)}] = \delta_{kl}$.
This establishes Eq.~\eqref{eq: exp value eigenstates occupation}. \\

This protocol can be further generalized to engineer a tailored momentum-space population distribution in the presence of multiple internal degrees of freedom within a unit cell. Interferences between them, which may generate additional structure in the momentum-space population, can be suppressed by assigning random phases. 

Indeed, in translationally invariant systems, it is natural to describe the steady state in momentum space. This allows one to analyze how the pumping profile distributes photons across Bloch bands and momenta. Taking the Fourier transform of Eq.~\eqref{eq: reduced Lindblad}, we obtain
\begin{align}
\left((\omega_0 + i\gamma)\delta_{\sigma \kappa} -H_{\sigma \kappa}(\vec k)\right) c_\kappa (\vec k) =f_\sigma(\vec k) \, ,
\end{align}
where $c_\kappa(\vec k)$ are the steady-state photon modes in momentum space, and $\sigma,\kappa$ label internal degrees of freedom within the unit cell. 

Here we denote the components of the photon field projected onto the Bloch basis by $u_\beta(\vec k) = P^\dg_{\beta \sigma}(\vec k) c_\sigma(\vec k)$, with $\beta$ the band index of energy $\varepsilon_\beta(\vec k)$, and $P_{\sigma \beta} = (e_\beta)_\sigma$ diagonalizes $H(\vec k)$. Hence we obtain
\begin{align}
|u_\beta(\vec k)|^2 = \sum_{\sigma \kappa } \frac{(e^*_\beta(\vec k))_\sigma f_\sigma(\vec k) \, (e_\beta(\vec k))_\kappa f^*_\kappa(\vec k)}{(\omega_0 - \varepsilon_\beta(\vec k))^2 + \gamma^2} \, .
\end{align}

In general, the occupation of the Bloch bands depends on the overlap between the Fourier transform of the pumping profile and the Bloch eigenvectors. 
However, if the pumping field components acquires independent and uniformly distributed random phases between the different internal degrees of freedom, such that $f_\sigma(\vec k) = f(\vec k) e^{i\theta_\sigma}$, the cross terms vanish after averaging. Namely, one then obtains
\begin{align}
\mathbb E[|u_\beta(\vec k)|^2] = \frac{|f(\vec k)|^2}{(\omega_0 - \varepsilon_\beta(\vec k))^2 + \gamma^2} \, .
\end{align}
Thus, the random phases suppress the Bloch eigenvector structures, such that it produces a population directly controlled by the pumping field envelope $|f(\vec k)|^2$ in momentum, whereas the band populations are filtered in energy by the Lorentzian distribution centered at the frequency $\omega_0$.

\section{From the coarse-grained energy-resolved \St response to the Chern number}

Equation~\eqref{eq: convolved Streda} 
provides a coarse-grained energy-resolved \St marker centered at $\omega_0$ with a resolution set by the loss rate $\gamma$. We now describe how the Chern number of a target Bloch band can be reconstructed from this coarse-grained \St marker. \\

We consider performing the DD protocol for $N_\omega$ different values of the pump frequency $\omega_0$, chosen such that the intervals $[\omega_0 - \delta \omega_0, \omega_0 + \delta \omega_0]$ associated with the $N_\omega$ values of $\omega_0$ partition the entire bandwidth of a target Bloch band. As motivated in the main text, we set $\delta \omega_0 / \gamma \equiv \zeta \sim O(1)$. For each $\omega_0$, we have [Eq.~\eqref{eq: convolved Streda}]  
\begin{align}
    &\mathbb E \left[ \Phi_0\frac{\delta n(\omega_0)}{\delta B} \right] = \Phi_0 \int \frac{\delta \rho_A(\omega)}{\delta B} \frac{|f|^2}{(\omega - \omega_0)^2 + \gamma^2} d\omega \notag \\ 
    &\qquad \qquad \, = \Phi_0 \frac{\pi |f|^2}{\gamma} \int \frac{\delta \rho_A(\omega)}{\delta B} \frac{\gamma / \pi }{(\omega - \omega_0)^2 + \gamma^2} d\omega \, ,
\end{align}
such that summing all the responses results in 
\begin{align}
    &\sum_{\omega_0} \mathbb E \left[ \Phi_0\frac{\delta n(\omega_0)}{\delta B} \right] = \sum_{\omega_0} \Phi_0 \frac{\pi |f|^2}{\zeta \, \gamma^2} \times \notag \\ 
    & \qquad \qquad  \times \left( \int \frac{\delta \rho_A(\omega)}{\delta B} \frac{\gamma/\pi}{(\omega - \omega_0)^2 + \gamma^2} d\omega \right) \delta \omega_0 \, , \label{eq: sum Streda}
\end{align}
where we introduced the parameter $\zeta\!=\!\delta \omega_0 / \gamma$.

We notice that the kernel in the integral, $\frac{\gamma/\pi}{(\omega - \omega_0)^2 + \gamma^2}$, converges to the Dirac distribution $\delta(\omega - \omega_0)$ in the limit $\gamma \to 0$. In order to take this limit in Eq.~\eqref{eq: sum Streda}, one should work in a regime where the prefactor $\pi |f|^2 / \zeta \, \gamma^2$ remains constant as $\gamma \to 0$. This is achieved by fixing the ratio $|f|^2/\gamma^2 \equiv \nu \sim O(1)$, where we introduced the ``filling factor" $\nu$. Then, upon replacing the sum by an integral $\sum_{\omega_0} \delta \omega_0 \to \int d\omega_0$ on the RHS of Eq.~\eqref{eq: sum Streda}, we obtain 
\begin{align}
    \sum_{\omega_0} \mathbb E \left[ \Phi_0\frac{\delta n(\omega_0)}{\delta B} \right] &= \frac{\pi \, \nu}{\zeta} \,  \Phi_0 \int_{\mathrm{band}} \frac{\delta \rho_A (\omega_0)}{\delta B} d\omega_0 \notag \\ 
    &=\frac{\pi \, \nu}{\zeta}\, \mathcal C \, ,
\end{align}
which recovers the Chern number of the occupied Bloch band up to the scaling factor $\frac{\pi \, \nu}{\zeta}$. To fix this factor to unity, and thus directly extract the Chern number, one can further set $f = \gamma$ (i.e. $\nu = 1$), and $\delta \omega_0= \pi \gamma$ (i.e. $\zeta = \pi$). 

\section{Flat band regime} 
Considering the DD protocol with the uniform random pump introduced in the main text [Eq.~\eqref{eq: uniform pump random}], one sees that Eq.~\eqref{eq: exp value eigenstates occupation} naturally identifies the flat‑band limit as a particularly relevant regime. Indeed, when the band dispersion is negligible, \textit{i.e.} $\varepsilon_\lambda \simeq \mathrm{const}$, and the loss rate $\gamma$ exceeds the band width, \textit{i.e.}, $\Delta_w \ll \gamma \ll \Delta_g$~\cite{Ozawa_2014}, the steady state leads to an approximately uniform population of the band eigenstates, $\mathbb E[|c_\lambda|^2] \approx |f|^2/\gamma^2 \equiv \nu$, for a pump frequency resonant with the flat-band energy. For such a uniform occupation, the density response directly reflects the quantized \St response [Eq.~\eqref{eq: streda thermod}], up to a multiplicative filling factor. Indeed, this result follows from integrating the energy-resolved \St response [Eq.~\eqref{eq: energy resolved Streda}] over the target band, assuming a uniform filling $\nu$, which yields a quantized response scaled by $\nu$. In the following, we set $f = \gamma$, such that $\nu =1$.

By evaluating the steady-state bulk photon density $n$ obtained from the DD protocol, both in the presence and in the absence of a small ``magnetic" perturbation, and repeating the procedure over $D$ independent realizations of the random drive, the resulting averaged density variation is then expected to obey $\Phi_0 \, \Bigl \langle \frac{\delta n}{\delta B} \Bigr \rangle_D \approx \nu \, \mathcal C$, where $\mathcal C$ is the Chern number of the band. 

This prediction can be understood from the perspective of spectral flow in systems with open boundaries, where the net number of transferred edge states into bulk states (or vice-versa) is fixed by the topological invariant associated with the gap  \cite{Asboth_2017, PeraltaGavensky_2025}. In practice, this process can be captured through a direct counting of states up to a spectral gap, which is equivalently obtained by integrating the DOS over the target band. We therefore introduce the following relation, which will serve as a benchmark to validate the spectral flow picture in open boundaries systems: 
\begin{align}
\mathcal C = \Phi_0 \, \frac{\delta N_s}{\delta \Phi}
= \Phi_0 \, \frac{\delta n_s}{\delta B}\, , \label{eq: DOS streda}
\end{align}
where $N_s$ counts the total number of eigenstates below a spectral gap, and $n_s\!=\!N_s/V$ with $V$ the system's volume. This expression [Eq.~\eqref{eq: DOS streda}] coincides with the \St response of a fermionic insulating system at zero temperature with a chemical potential inside the gap.

\begin{figure}[t]
    \centering
    \includegraphics[scale=0.42]{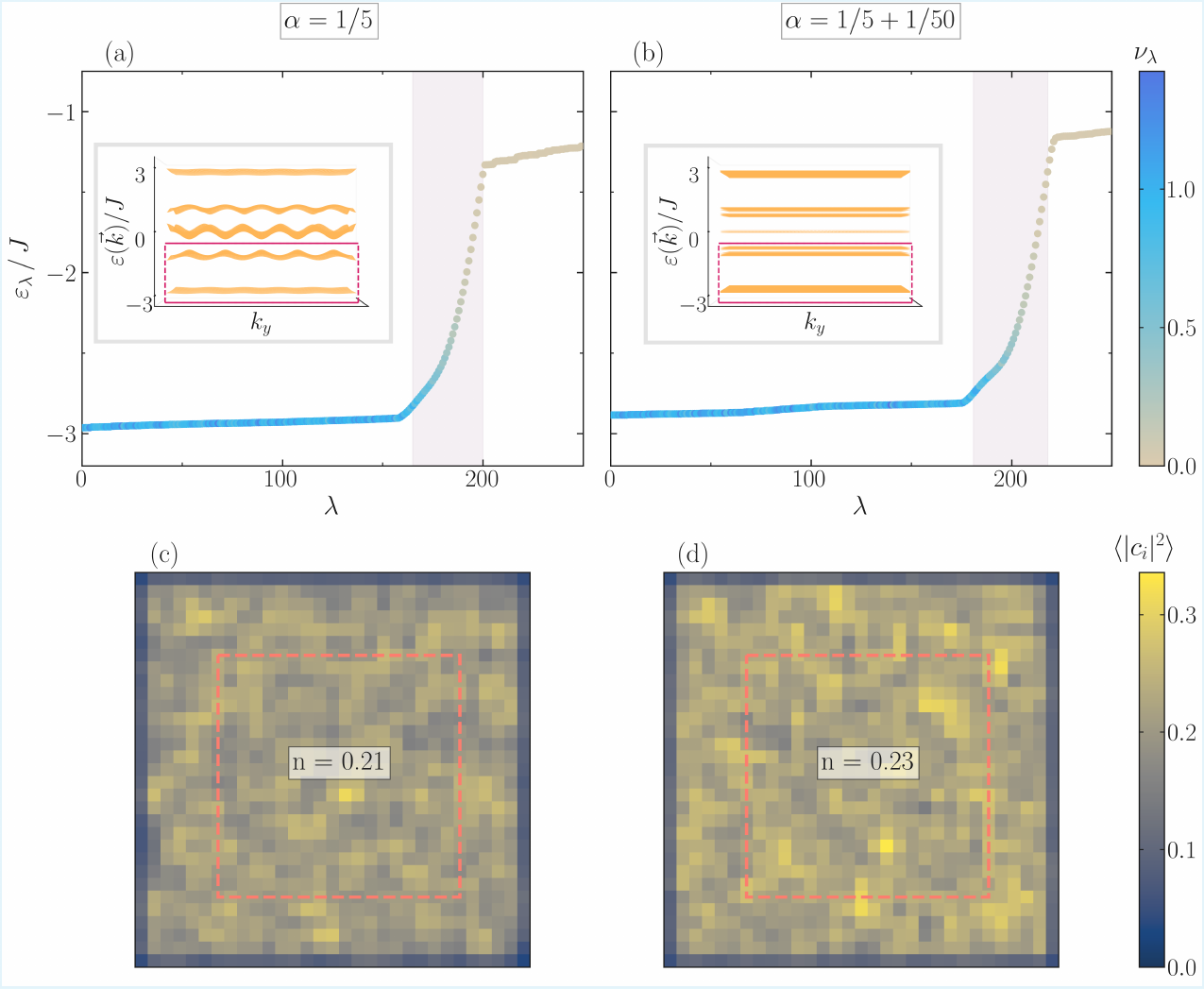}
    \caption{Energy spectrum of the Harper-Hofstadter model on a finite sample of $31 \times 31$ unit cells with open boundary conditions, zoomed in on the lowest band up to the onset of the second band, for (a) $\alpha = 1/5$ and (b) $\alpha = 1/5 + 1/50$. The respective insets show the energy band spectra with periodic boundary conditions, with an emphasis on the first two bands. The horizontal axis $\lambda$ in panels (a) and (b) labels the eigenvalues. The shaded region highlights the edge modes, defined here as those for which more than 92\% of the particle occupation is localized within the first two layers of unit cells adjacent to the boundary of the square lattice. Each mode is colored according to the expectation value of the eigenstate occupation $\nu_\lambda$ obtained from a simulation of the DD protocol, averaged over $D=50$ realizations. The colorbar ranges from blue ($\nu_\lambda \approx 1$) to dark yellow ($\nu_\lambda \approx 0$). We use $\gamma/J = 0.3 = f/J$, and $\omega_0/J$ is set to $-2.93$ in (a) and $-2.85$ in (b). Panels (c) and (d) display the corresponding photon-amplitude distribution in the lattice $\langle |c_i|^2\rangle$. The inset shows the photon density averaged over the subregion outlined by the red dashed box, from which we extract $\langle \Phi_0 \, \delta n / \delta B \rangle_{50} = 0.95$ in the simulation with $D=50$ realizations.}. 
    \label{fig:protocol flat band}
\end{figure}
  
We now present the numerical validations of the DD scheme on the Harper-Hofstadter model \cite{Harper_1955, Hofstadter_1976}, which provides a simple lattice realization of topological flat bands. The system is defined on a $31 \times 31$ square lattice with nearest-neighbor hopping $J$ and subjected to a perpendicular constant magnetic field $\boldsymbol B = B \, \h z$, which we chose to implement with a Landau gauge $A = B(0,x,0)$. We parametrize the field through the dimensionless flux $\alpha$, introduced via the magnetic flux per unit cell $\Phi_{cell} = \alpha \Phi_0 = 2\pi \alpha$ (with $\hbar = 1 = e$).

The numerical results are presented in Fig.~\ref{fig:protocol flat band}. Figures~\ref{fig:protocol flat band}(a) and (b) display the energy spectrum of the Harper-Hofstadter model, where the eigenvalues $\varepsilon_\lambda$ are labeled by $\lambda$ up to the second band. We set $\alpha = 1/5$ (a), placing the unperturbed system in a regime where the lowest band is nearly flat and has a non-trivial Chern number $\mathcal C=1$, and is well separated from the others.
The magnetic perturbation of amplitude $\delta \alpha = 1/50$ then modifies the spectrum, as shown in Fig.~\ref{fig:protocol flat band}(b). 

As a benchmark, we have verified that a direct evaluation of the variation of the density of states of the model, obtained from Eq.~\eqref{eq: DOS streda}, correctly captures the spectral flow with $\mathcal C = 1$, which we now compare to the results obtained from the DD protocol. In Figs.~\ref{fig:protocol flat band}(a) and (b), we report the averaged eigenmodes population simulated over $D = 50$ realizations as a colormap. For the protocol to operate successfully, we set $\gamma/J = 0.3$ such that the loss rate satisfies $\Delta_w \ll \gamma \ll \Delta_g$. In the presence of a finite magnetic perturbation (here, $\delta \alpha\!=\!1/50$), the flatness ratio of the band is slightly reduced. As a result, while the pumping frequency is set to $\omega_0/J = -2.83$ for the unperturbed system [Fig.~\ref{fig:protocol flat band}(a)], we slightly re-adjusted the pump frequency at the value $\omega_0/J\!=\!-2.85$ for the perturbed case [Fig.~\ref{fig:protocol flat band}(b)], corresponding to resonance with the center of the lowest band: this choice ensures a uniform population of the lowest band in both cases, with $\nu = \langle |c_\lambda|^2\rangle_D \approx 1$ for all $\lambda$ in the bulk states belonging to the lowest band (blue), while the edge states are less populated (dark yellow). \\
Figures~\ref{fig:protocol flat band}(c) and (d) provide a map of the averaged steady-state photon amplitude in the lattice, associated with the eigenmodes population of Figs.~\ref{fig:protocol flat band}(a) and (b). One observes that the bulk becomes brighter as the magnetic perturbation is applied, whereas the edges remain essentially dark.
This directly reflects the spectral flow, with edge states acting as a reservoir for the bulk. From the averaged bulk density [Eq.~\eqref{eq: photon bulk density}] within the highlighted subregion (red dashed square of size $19 \times 19$ cells [Figs.~\ref{fig:protocol flat band}(c)-(d)]), we can then extract $\Phi_0 \, \langle \delta n / \delta B \rangle_{D}$.
Since the latter value fluctuates due to inherent randomness of the protocol, we repeat the simulation $N_{rep} = 20$ times for $D$ realizations. The \St response obtained from the corresponding statistical mean value is then $0.93$, with a standard deviation given by $\sigma_{dev} = 0.24$ for $D = 50$, in good agreement with the Chern number of the lowest band $\mathcal C = 1$.

We note that in the infinitesimal limit $\delta \alpha \to 0$, one recovers a quantized response using a single driving frequency $\omega_0$ for both the perturbed and the unperturbed regime. Finally, increasing the number of realizations brings the eigenmodes occupation closer to the ideal one of Eq.~\eqref{eq: exp value eigenstates occupation}: we verified that this improves the accuracy of the result, as it diminishes the standard deviation [Appendix]. 

\subsection{Standard deviation analysis}

\begin{figure}[h]
    \centering
    \includegraphics[scale=0.34]{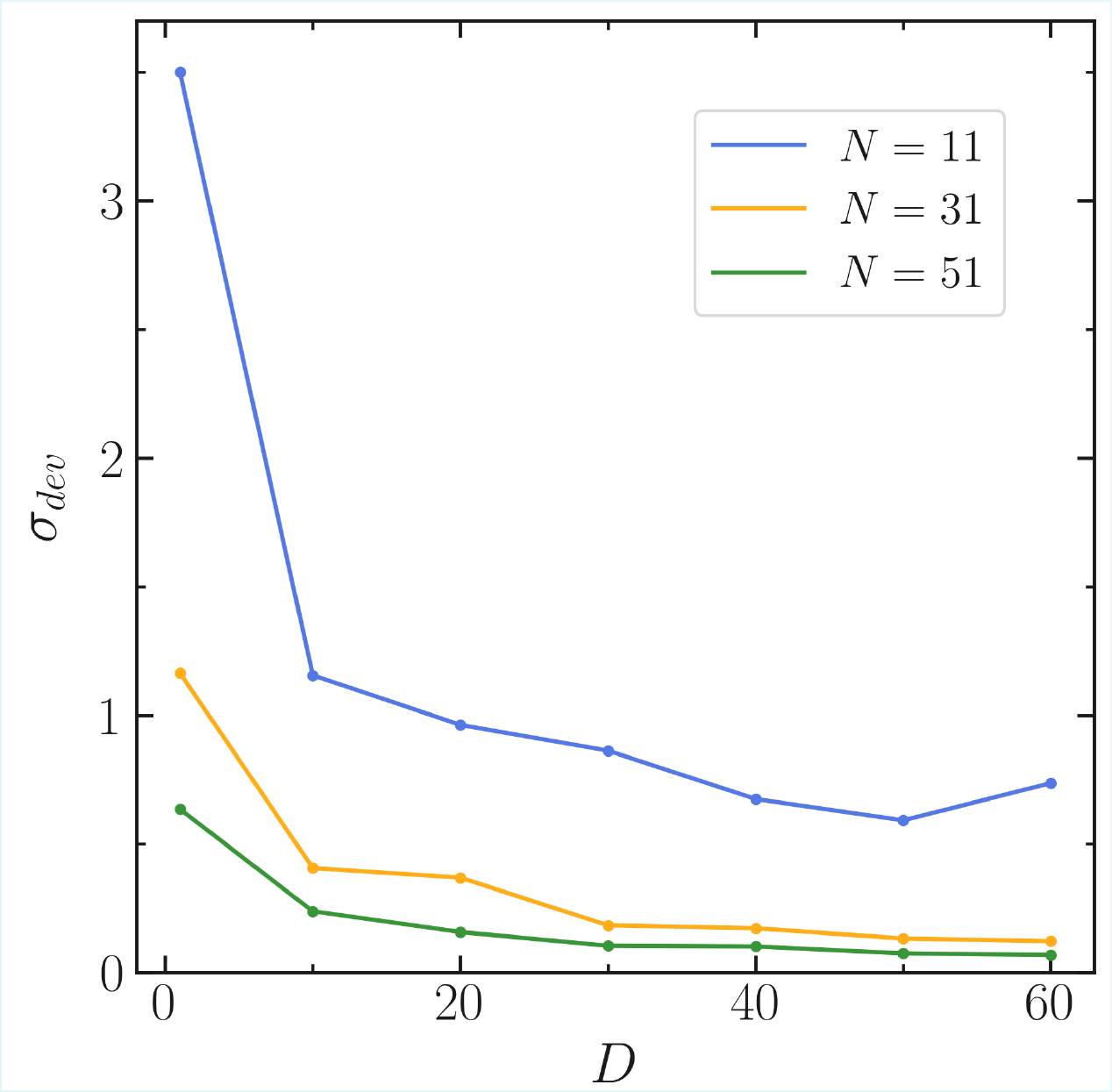}
    \caption{Standard deviation $\sigma_{dev}$, defined in Eq.~\eqref{eq: appendix sigma and mean value}, obtained with $N_{rep} = 20$, as a function of the number of realizations, for three different system sizes of the Harper-Hofstadter model at flux $\alpha = 1/5$: $N=11$ (blue), $N = 31$ (orange) and $N=51$ (green) and the bulk subregions over which the bulk density is averaged are respectively $N_a = 5\times 5$, $N_a = 19 \times 19$ and $N_a = 39 \times 39$. We use the same parameters $\omega_0$, $\gamma$, and magnetic perturbation $\delta \alpha$ as in Fig.~\ref{fig:protocol flat band}.}
    \label{fig: D_realizations}
\end{figure}

The DD scheme presented in the main text produces inherently random outcomes due to the random phases entering the pumping protocol. In the limit of an infinite number of realizations, one expects the steady-state population to converge to the ideal distribution given by Eq.~\eqref{eq: exp value eigenstates occupation}, and consequently, the extracted St\v{r}eda response to become free of fluctuations. Here, we examine how the standard deviation of the measured St\v{r}eda response decreases with the number of realizations $D$. \\ 

For a fixed number of realizations $D$, we repeat the simulation $N_{rep}$ times in order to determine a statistical mean value $\bar X$ and standard deviation $\sigma_{dev}$ of the dataset $\left\{ X_i = \bigl \langle \Phi_0 \frac{\delta n}{\delta B}\bigr \rangle_D^{(i)} \right\}$, where $X_i$ denotes the $i^{th}$ outcome obtained using $D$ realizations:
\begin{align}
\begin{cases}
\bar X = \frac{1}{N_{rep}}\sum_i^{N_{rep}} X_i \\
\sigma_{dev} = \sqrt{ \frac{1}{N_{rep}-1} \sum_{i=1}^{N_{rep}} (X_i - \bar X) ^2}\, . \label{eq: appendix sigma and mean value}
\end{cases}
\end{align}
The quantity $\sigma_{dev}$ therefore quantifies the residual fluctuations of the St\v{r}eda response when averaging over $D$ realizations of the random pumping. \\ 

Figure~\ref{fig: D_realizations} displays the standard deviation obtained following the procedure above as a function of the number of realizations, for the DD protocol applied to the Harper-Hofstadter model at flux $\alpha = 1/5$ for three different system sizes, $N \times N$ unit cells. We use the same parameters $\gamma$, $\omega_0$ and $\delta \alpha$ as in Fig.~\ref{fig:protocol flat band}. For all system sizes, $\sigma_{dev}$ decreases as $D$ increases, confirming that the protocol converges toward a response with high accuracy in the large-$D$ limit. \\
In addition, for a fixed number of realizations, increasing the system size reduces the fluctuations. This behavior reflects an averaging effect: for larger system sizes, the bulk density entering the St\v{r}eda response can involve an average over a larger number of states. As a result, the fluctuations are partially averaged out in larger systems. \\
In the dispersive regime, smaller loss rate $\gamma$ probe a smaller window of eigenstates and are therefore more sensitive to fluctuations. Achieving a stable signal in this case then requires larger system sizes and a greater number of realizations.

\section{Finite size scaling}

\begin{figure}[h]
    \centering
    \includegraphics[scale=0.4]{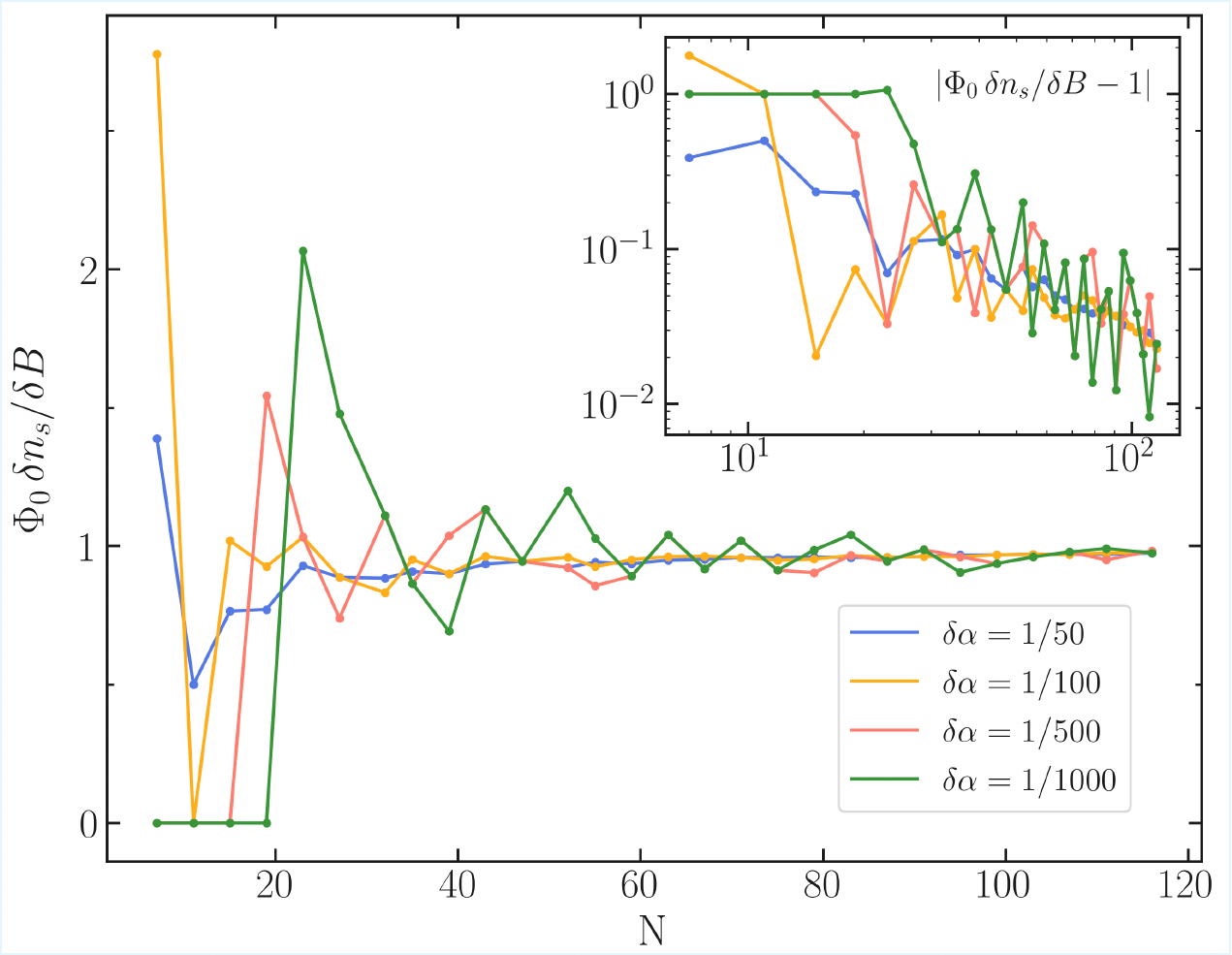}
    \caption{Finite-size scaling of the integrated St\v{r}eda response [Eq.~\eqref{eq: DOS streda}] up to the first spectral gap of the Harper-Hofstadter model with $\alpha = 1/5$, as a function of the system size. We display this quantity for four different strength of the magnetic perturbation: $\delta \alpha = 1/50$ (blue), $\delta \alpha = 1/100$ (orange), $\delta \alpha = 1/500$ (red) and $\delta \alpha = 1/1000$ (green). The inset displays $|\Phi_0 \frac{\delta n_s}{\delta B} -1|$, in log-log scale as a function of $N$, for the same values of $\delta \alpha$.}
    \label{fig: finitesize}
\end{figure}

In this section, we verify that the integrated St\v{r}eda response in finite systems converges toward the quantized value following a power-law behavior as the system size increases. The corresponding finite-size scaling is shown in Fig.~\ref{fig: finitesize}, where we plot the integrated St\v{r}eda response [Eq.~\eqref{eq: DOS streda}] as a function of the system size $N$ in the Harper-Hofstadter model with $\alpha = 1/5$, for several values of the magnetic perturbation $\delta \alpha$. The inset displays the absolute deviation from the quantized value, \textit{i.e.} $|\Phi_0 \frac{\delta n_s}{\delta B} -1|$, in log-log scale as a function of $N$. The resulting linear behavior indicates a power-law convergence toward the quantized limit. 

For a fixed perturbation strength, a minimal system size is required for the response to be accurately resolved. Below this size, the magnetic length associated with the perturbation becomes comparable to or larger than the system size, preventing a faithful probing of the bulk response. \\

\section{Finite field effect}

\begin{figure}[h]
    \centering
    \includegraphics[scale=0.43]{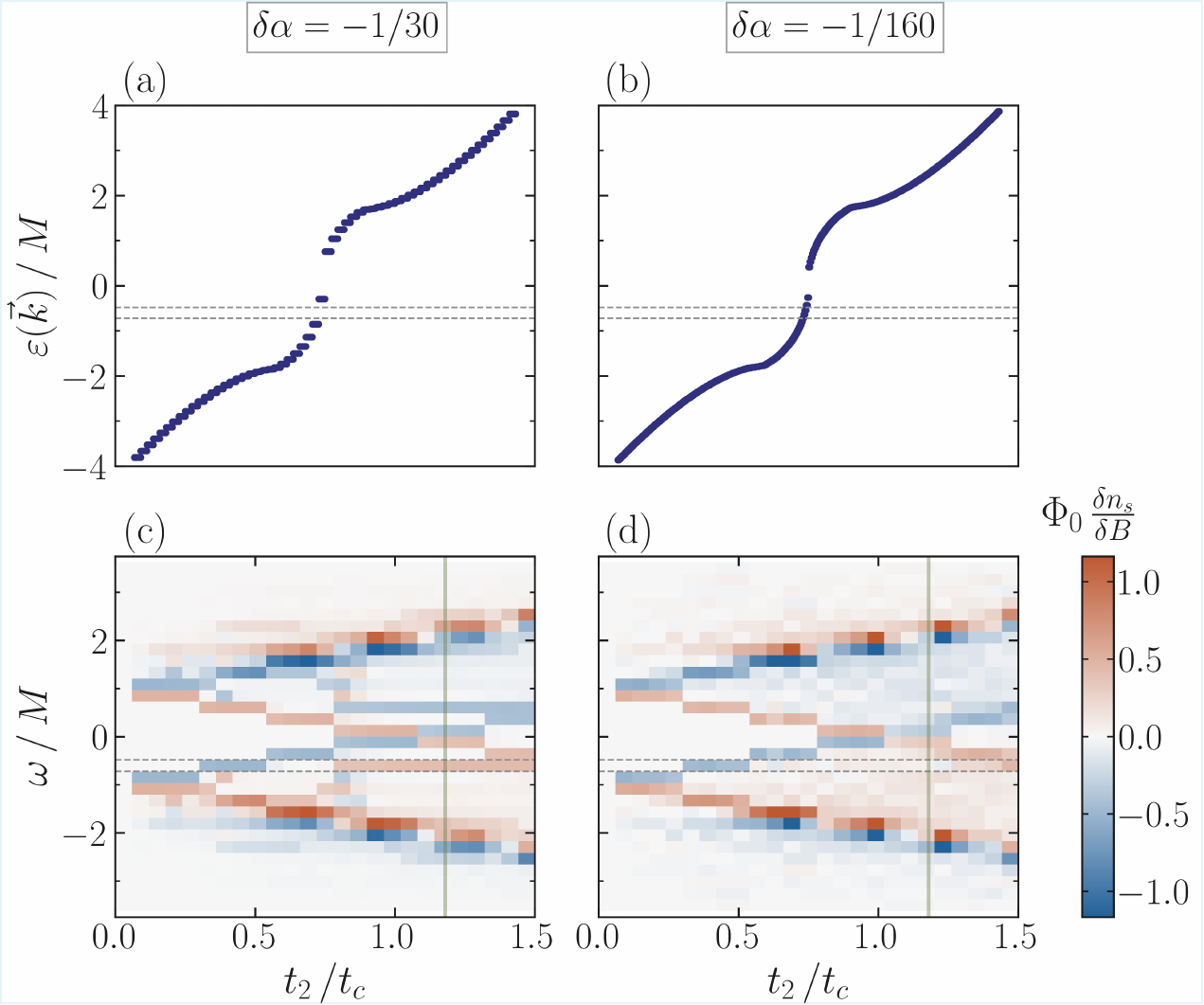}
    \caption{Finite magnetic-field effects in the energy-resolved St\v{r}eda response of the Haldane model. Panels (a) and (b) display the energy-resolved band spectra in momentum space at $t_2/t_c=1.2$ (vertical green lines in panels (c) and (d)), for $\delta\alpha=-1/30$ (a) and $\delta\alpha=-1/160$ (b). Panels (c) and (d) show the corresponding finite centered difference approximation of the magnetic response [Eq.~\eqref{eq: centered diff}], displayed using $N_\omega = 30$ energy-resolved pixels as a function of $t_2/t_c$, for $\delta\alpha=1/30$ (c) and $\delta\alpha=1/160$ (d). In all panels, we set $M/t_1 = 0.8$ and $\phi = \pi/2$.}
    \label{fig: finiteB}
\end{figure}

We now turn to finite magnetic perturbation effects in the energy-resolved St\v{r}eda response, which can produce singular behavior in the presence of a relativistic Landau-level structure near Dirac cones. Indeed, in this regime, energy levels shift rapidly with magnetic field, leading to a non-regular dependence of the density of states on $B$. We illustrate this effect on the Haldane model in Fig.~\ref{fig: finiteB}: in panels (c) and (d), we show the magnetic response of the density of states evaluated here with a finite centered difference
\begin{align}
    \Phi_0 \int_{\omega_0 - \delta \omega_0}^{\omega_0 + \delta \omega_0} \frac{\rho(\omega \, ; B) - \rho(\omega \, ; -B)}{2\, \delta B} d\omega \, , \label{eq: centered diff}
\end{align}
integrated over intervals $[\omega_0 - \delta \omega_0, \omega_0 + \delta \omega_0]$ for $N_\omega = 30$ different values of $\omega_0$, with $\delta \omega_0$ chosen to partition the entire spectrum. An additional feature is visible in panel (c) for intermediate values of $t_2 / t_1$ and for energies close to the gap. For sufficiently large $\delta \alpha$, the discrete Landau-level structure of the bulk spectrum implies that bulk states may lie inside a given energy window for $\delta \alpha$ while being absent for $-\delta \alpha$, producing a finite contribution to the DOS variation. 

This mechanism is illustrated in panels (a) and (b), which show the energy-resolved band spectrum of the Haldane model obtained in momentum space for magnetic perturbations $\delta \alpha = -1/30$ (a) and $\delta \alpha = -1/160$ (b), at $M/t_1 = 0.8$ and $t_2/t_c = 1.2$ (highlighted by the green vertical lines in panels (c) and (d)). In panel (a), no bulk states are present in the considered energy window, while states appear for $\delta \alpha = 1/30$ (not shown), resulting in a finite density variation. For smaller perturbations $\delta \alpha = \pm 1/160$ [panel (b)], bulk states populate all energy windows up to the gap, smoothing the dependence of the density of states on the magnetic field. The additional signal is therefore strongly suppressed in panel (d), confirming that the feature observed in panel (c) reflects a finite-field spectral effect rather than the infinitesimal-field St\v{r}eda response.

\section{The local Chern marker in disordered systems}

\begin{figure}[h]
    \centering
    \includegraphics[scale=0.4]{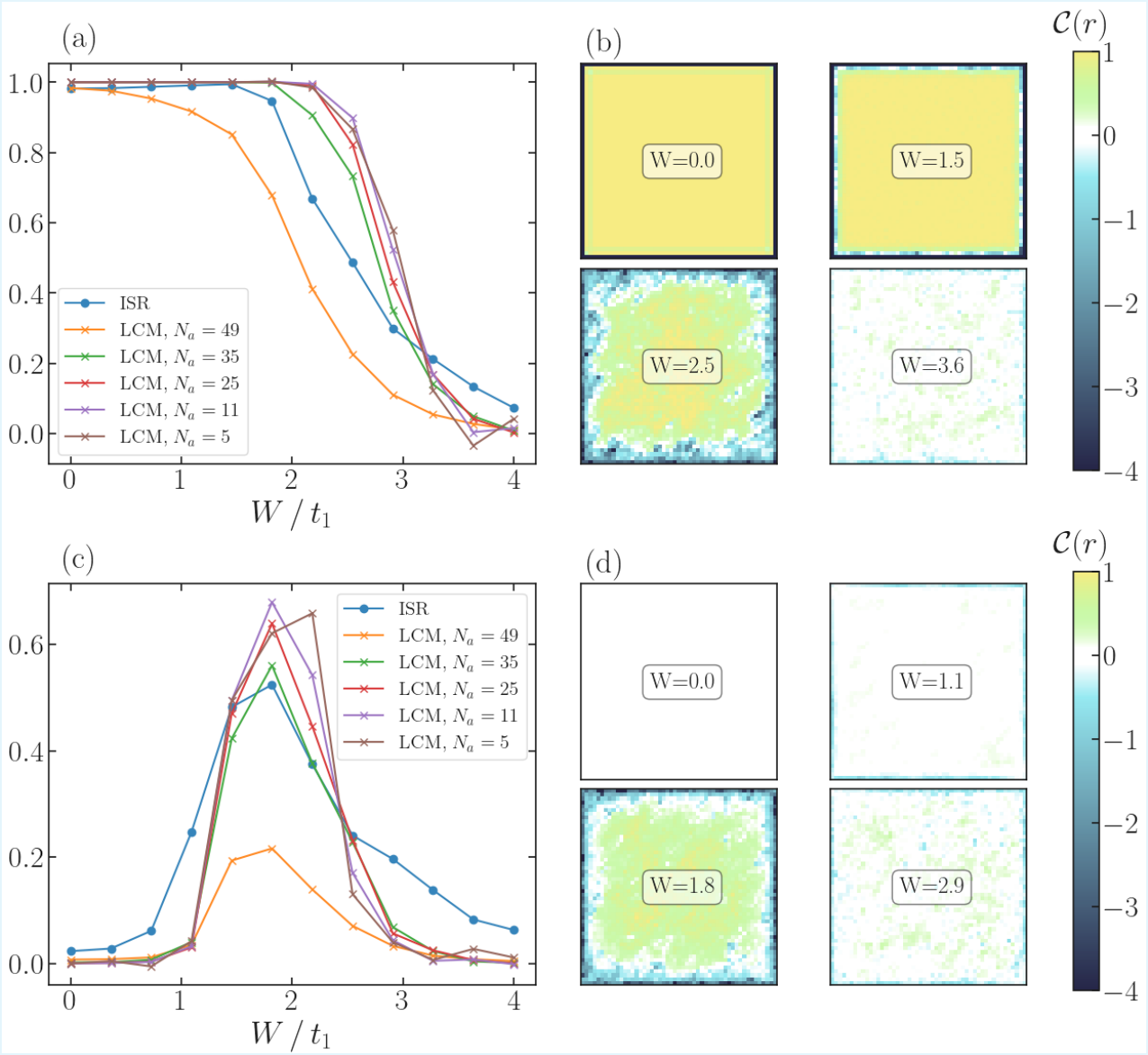}
    \caption{Evolution of the LCM [Eq.~\eqref{eq: def LCM}] and the ISR as a function of disorder strength $W$, for the Haldane model of size $51\times 51$ with $\phi = \pi/2$. (a)-(b) The system is initialized in a topologically non-trivial regime with $M/t_1 = 0$ and $t_2/t_1 = 0.2$. (a) We present the LCM averaged over bulk subregions of varying size $N_a$ (cross marks), compared to the ISR obtained from the bulk DOS \eqref{eq: bulk proj DOS} with a bulk of size $47 \times 47$ (blue dots). (b) Real-space maps of $\mathcal{C}(\textbf{r})$ at selected disorder strengths $W/t_1 \in \{0.0, 1.5, 2.5, 3.6\}$. (c)-(d) The system is initialized in a trivial regime with $M/t_1 = 0.8$ and $t_2 = 0.8\, t_c$. (c) Same as (a). (d) Real-space maps of $\mathcal{C}(\textbf{r})$ at selected disorder strengths $W/t_1 \in \{0.0, 1.1, 1.8, 2.9\}$. In all real-space maps, the colorbar is clipped from below at $-4$. Results are averaged over $20$ disorder realizations.}
    \label{fig: LCM vs ISR}
\end{figure}

Topology in disordered systems can be well identified in real space through the so-called local Chern marker (LCM) \cite{Bianco_Resta_2011, Lee_2022, Prodan2011, Kovalenka_2026}, which is defined as
\begin{align}
    \mathcal C(\textbf{r}) = -\frac{4\pi }{A_{cell}} \mathrm{Im} \mathrm{Tr}_{cell}[\h P \h X\h Q \h Y \h P] \, , \label{eq: def LCM}
\end{align}
where \textbf{r} denotes the position of a unit cell, $A_{cell}$ is the area of a unit cell, and $\h Q = \h I - \h P$ with $\h P$ a projector onto energy states below a reference energy lying in a gap, which will correspond here to $\omega = 0$.
In the main text, we introduced the \St response as a practical tool to probe topology. In the absence of disorder and in a zero-temperature fermionic gapped system, the integrated \St response (ISR) up to a spectral gap is known to reproduce the Chern number of the occupied bands, which can equivalently be obtained from the local Chern marker averaged over bulk states.
As disorder is introduced, we have seen that the ISR still provides signatures of topology; we therefore benchmark our results against the LCM to corroborate this topological interpretation.

Figure~\ref{fig: LCM vs ISR} displays the evolution of the LCM [Eq.~\eqref{eq: def LCM}] and the ISR $\Phi_0 \int^0\delta \rho_A (\omega)/ \delta B \, d\omega $ as disorder is introduced in the Haldane model of size $51 \times 51$.
In Figs.~\ref{fig: LCM vs ISR}(a)-(b), the system is initialized in a topologically non-trivial regime. In Fig.~\ref{fig: LCM vs ISR}(a), we display the evolution of the LCM averaged over a subregion of size $N_a$, which shows good agreement with the ISR, with a sharper transition at a critical disorder strength $W_c$ toward a vanishing LCM as the subregion is taken smaller. This observation is further supported by Fig.~\ref{fig: LCM vs ISR}(b): while in the clean limit ($W=0$) the LCM is uniformly quantized in the bulk, with edge states contributing negative values sharply localized along the edge, the gap progressively closes as disorder is increased and bulk quantization becomes increasingly affected by edge contributions extending into the bulk.

In Figs.~\ref{fig: LCM vs ISR}(c)-(d), we repeat the same analysis with the system initialized in a trivial regime of the Haldane model, while still breaking time-reversal symmetry.
In this regime, we identified in the main text an intermediate region where edge states emerge and the ISR increases accordingly. We use the LCM to validate the topological nature of this behavior. Figure~\ref{fig: LCM vs ISR}(c) shows again that the LCM averaged over a subregion of size $N_a$ follows a trend accurately captured by the ISR. In Fig.~\ref{fig: LCM vs ISR}(d), we observe that starting from a uniformly vanishing LCM in the clean limit ($W=0$), the system reaches a regime where, deep in the bulk, $\mathcal C(\textbf{r})$ attains a quantized value. Furthermore, we have verified that averaging the LCM over a two-layer-thick edge yields a large finite value, which can also be interpreted as a topological signature \cite{Kovalenka_2026}. However, finite-size effects can be significant since the gap is small, and as a result both the bulk-averaged LCM and the ISR are strongly affected by extended edge contributions, causing the markers to deviate from $1$: we further validate this picture by performing a finite-size scaling analysis in the next section.

\section{Finite size effects in disordered systems}

\begin{figure}[b]
    \centering
    \includegraphics[width=\columnwidth]{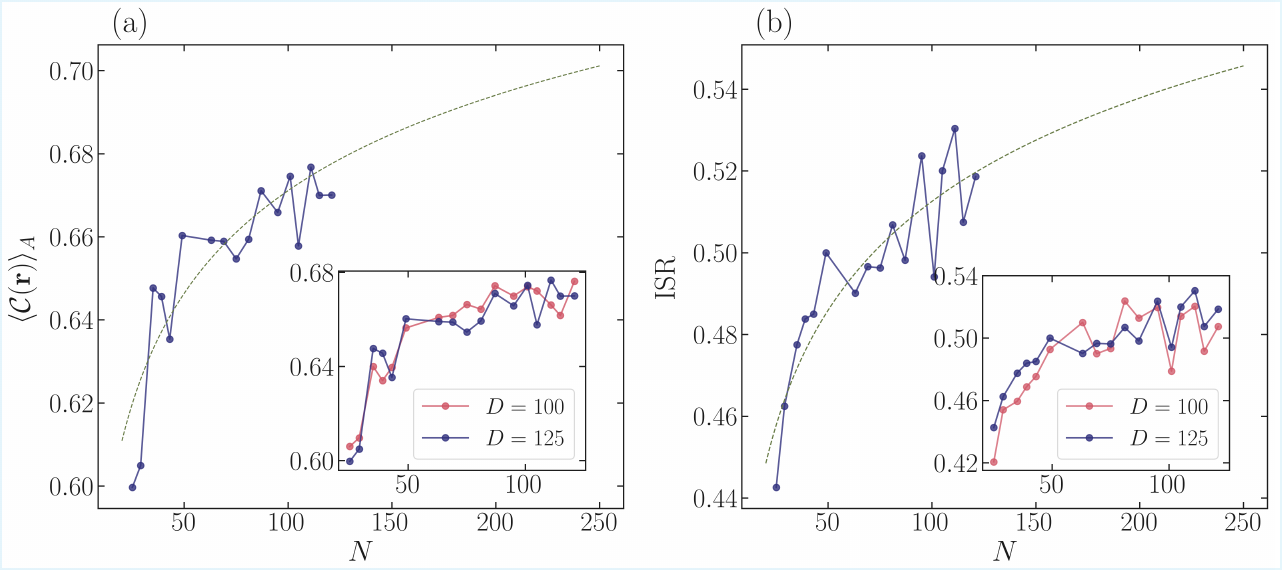}
    \caption{Finite-size scaling of the bulk averaged LCM and the ISR at fixed disorder strength $W/t_1 = 1.76$, for the Haldane 
model with $M/t_1 = 0.8$, $t_2 = 0.8\, t_c$, and $\phi = \pi/2$. 
(a) LCM $\langle \mathcal{C}(\textbf{r}) \rangle_A$ averaged over a bulk 
subregion comprising approximately $75\%$ of the total sample, as a function of system 
size $N$, computed via KPM with $M_\mathrm{LCM} = 300$ moments. 
(b) ISR as a function of $N$, computed via KPM with $M_\mathrm{ISR} = 1200$ 
moments, using a finite magnetic perturbation $\delta\alpha = 1/40$. 
In both panels, the main curve is obtained with $D = 125$ disorder realizations and the 
dashed line is a power-law fit. Insets present the same results with both $D = 100$ and $D = 125$ realizations, showing convergence with respect to the number of disorder realizations.}
\label{fig: finite size TAI}
\end{figure}
We focus on the regime of Figs.~\ref{fig: streda disorder}(b) and (d), where the system exhibits topologically non-trivial behavior at strong disorder despite being initialized in a trivial regime.
Having shown above that the integrated \St response accurately tracks the evolution of the local Chern marker over a finite bulk subregion across topological transitions, we now perform a finite-size scaling analysis of these quantities in this regime at a fixed disorder strength $W/t_1 = 1.76$ to support the interpretation of a topological Anderson insulator.

Figures.~\ref{fig: finite size TAI}(a)-(b) display the LCM averaged over a subregion comprising a fixed fraction of the entire sample (approximately $75\%$), and the ISR, both as a function of system size $N$. The kernel polynomials method (KPM) \cite{Fehske_2006} has been used to compute the LCM [Eq.~\eqref{eq: def LCM}] and the density of states of the entire system $\rho (\omega)$, and the results are in good agreement with a power-law fit. Each curve is obtained by averaging over $D=100$ and $D=125$ disorder realizations, shown together as insets in each panel to demonstrate convergence with respect to the number of realizations.
These results indicate that as the system size increases, a larger fraction of bulk states reaches a quantized LCM. Edge contributions are consequently reduced, and both markers are therefore expected to converge to a quantized value in the thermodynamic limit: a hallmark of a topological Anderson insulator regime and the opening of a mobility gap. 

\end{document}